\documentclass[12pt,manuscript]{aastex}

\usepackage{times,emulateapj5}

\usepackage{psfig}

\def\simless{\mathbin{\lower 3pt\hbox
     {$\rlap{\raise 5pt\hbox{$\char'074$}}\mathchar"7218$}}}   
\def\simmore{\mathbin{\lower 3pt\hbox
     {$\rlap{\raise 5pt\hbox{$\char'076$}}\mathchar"7218$}}}   

\received{}
\accepted{}
\slugcomment{Submitted to Ap.J. Letters, June 16, 2000}

\shorttitle{M. M\'endez et al.}
\shortauthors{Kilohertz QPOs in 4U 1728--34, 4U 1608--52, and Aql X-1}

\begin{document}

\title{The Amplitude of the Kilohertz Quasi-periodic Oscillations
in 4U 1728--34, 4U 1608--52, and Aql X-1, as a Function of X-ray
Intensity}

\author{Mariano M\'endez\altaffilmark{1,2},
        Michiel van der Klis,
        Eric C. Ford
}
\affil{Astronomical Institute ``Anton Pannekoek'',
       University of Amsterdam and Center for High-Energy Astrophysics,
       Kruislaan 403, NL-1098 SJ Amsterdam, the Netherlands}

\altaffiltext{1}{Also Facultad de Ciencias Astron\'omicas y Geof\'{\i}sicas, 
       Universidad Nacional de La Plata, Paseo del Bosque S/N, 
       1900 La Plata, Argentina}
\altaffiltext{2}{Present address: SRON, Laboratory for
       Space Research, Sorbonnelaan 2, NL-3584 CA Utrecht, the
       Netherlands}

\begin{abstract}

We study the kilohertz quasi-periodic oscillations (kHz QPOs) in the
low-mass X-ray binaries 4U 1728--34, 4U 1608--52, and Aql X-1. Each
source traces out a set of nearly parallel tracks in a frequency vs.
X-ray count rate diagram. We find that between two of these tracks, for
similar QPO frequency, the $2-60$ keV source count rate can differ by
up to a factor of $\sim 4$, whereas at the same time the rms amplitude
of the kHz QPOs is only a factor of $\sim 1.1$ different. We also find
that, for 4U 1608--52 and Aql X-1, the rms spectrum of the kHz QPOs
does not depend upon which track the source occupies in the frequency
vs. X-ray count rate diagram. Our results for 4U 1728--34, 4U 1608--52,
and Aql X-1 are inconsistent with simple ``extra source of X-rays''
scenarios for the parallel tracks, such as those in which the properties
of the kHz QPOs are only determined by the mass accretion rate through 
the disk, whereas X-ray count rate also depends upon other sources of 
energy that do not affect the QPOs.

\end{abstract}


\section{Introduction} \label{introduction}

It is thought that the kilohertz quasi-periodic oscillations (kHz QPOs)
in low-mass X-ray binaries reflect the motion of matter in orbit at
some preferred radius in the accretion disk around the neutron star
\citep*{miller98, stella99, osherovich99, cui2000, campana2000}. In
most sources, two simultaneous kHz QPOs have been detected, with
frequencies between 300 and 1300 Hz. In a given source, the frequency
of the QPOs can shift by a few hundred Hz, apparently as a function of
$\dot M$, the rate at which mass is accreted onto the neutron star
\citep{miller98}. Yet, it is well known that the relation between the
kHz QPO frequencies and X-ray count rate (or X-ray flux), which is
assumed to be a good indicator of $\dot M$, is complex
\citep{ford0614_1, zhangAqlx1, mendez_1608_3}. In a frequency vs.
intensity ($\nu - I_{\rm X}$) diagram a source displays a set of almost
parallel tracks. On time scales of a day or less the source moves along
one of these tracks with QPO frequency and X-ray count rate positively
correlated, whereas on long time scales, in observations separated by a
few days the source occupies different tracks in this diagram
\citep[see, e.g., Fig. 2 in] []{mendez_1608_3}.

On time scales of a few days, QPO frequency correlates much better with
the X-ray spectral properties of the source than with count rate. In 4U
0614+09 it was found that the frequencies of the kHz QPOs correlate
better to the flux of the soft (``blackbody'') component in the energy
spectrum than to the total X-ray flux \citep{ford0614_2}, whereas a
good correlation of QPO frequency with photon index of the power-law
component in the energy spectrum has been found in 4U 0614+09 and 4U
1608--52 \citep{kaaret0614_1608}. The frequencies of the kHz QPOs also
correlate much better with the position of the source in an X-ray
color-color diagram than with count rate \citep{wijnands17+2,
wijnands5-1, jonker340+0_1, mendez_1608_3, mendez_1728, kaaret1820,
mendez_texas, reigAqlx1, steve0614, disalvo_1728}. This correlation
extends to other timing properties as well. The power spectra of these
sources often show a broad-band noise component that is roughly flat
below a break frequency at $\nu_{\rm b} \sim 1 - 10$ Hz and decreases
above $\nu_{\rm b}$, and a low-frequency QPO at $\sim 10 - 50$ Hz.
Studies with EXOSAT and Ginga already demonstrated that the rms
amplitude and break frequency of the broad band noise component, and
the frequency and rms amplitude of the low-frequency QPO depend
monotonically upon the position of the source on the color-color
diagram, and correlate much better to colors than to count rate
\citep[see][for a review]{vdk95}. Studies with the {\em Rossi X-ray
Timing Explorer (RXTE)} confirmed this across a much wider range of
sources \citep*{wijnands17+2, wijnandscygx2, wijnands5-1,
jonker340+0_1, markwardt, boirin_1916, jonker340+0_2, disalvo_1728,
bloser1820}, and also showed that both $\nu_{\rm b}$ and the frequency
of the low-frequency QPO are strongly correlated to the frequency of
the kHz QPOs \citep*{stella_lt,ford&vdk98, pbk99, disalvo_1728} and to
each other \citep{wk99}.

These correlations have reinforced the idea  \citep{hasinger89} that
there is one single parameter, inferred to be $\dot M$, which governs
all the timing and spectral properties of these sources; X-ray flux,
which can change more or less independently of the other parameters, is
the exception \citep[see also][]{vdk90,mendez_1608_3}.

In principle, the disparate behavior of X-ray flux and $\dot M$ could
be understood if mass flows onto the neutron star through two
independent channels \citep*[e.g.,][]{ghosh79, fortner89}. Part of the
mass accretion onto the star comes from the disk flow, $\dot M_{\rm
D}$, whereas the rest of the mass flux onto the star comes from a
radial flow, $\dot M_{\rm R}$. The timing and spectral properties are
governed by $\dot M_{\rm D}$, whereas the X-ray
\psfig{file=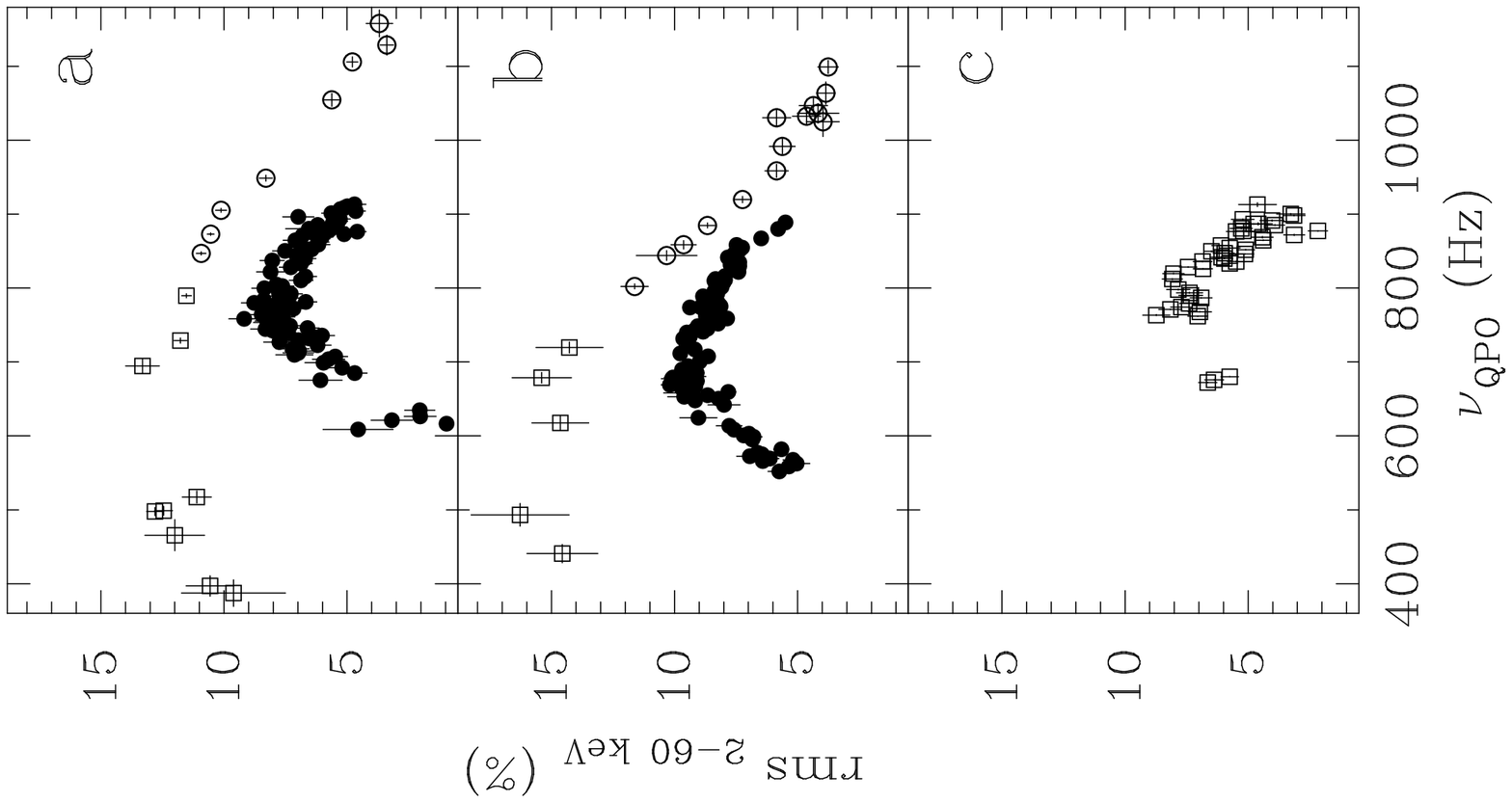,angle=-90,width=7cm}
\figcaption[fig1.ps]{The amplitude of the kHz QPOs as a function of
their frequency for (a) 4U 1728--34, (b) 4U 1608--52, and (c) Aql X-1.
For 4U 1728--34 and 4U 1608--52 the filled and open circles show the
relation for the lower and the upper kHz QPO, respectively, and open
squares indicate measurements in which we only detect a single QPO.
For Aql X-1 the open squares show the relation for the only kHz QPO
so far detected. The rms amplitudes are for the full PCA energy band.
\label{fig_rms_freq}}

\vspace{0.2cm}
\noindent
luminosity is
proportional to $\dot M_{\rm D} + \dot M_{\rm R}$ \citep{wijnands96,
kaaret0614_1608, mendez_1608_3}. In this scenario, the frequencies of
the kHz QPOs are linked to $\dot M_{\rm D}$, but
because $\dot M_{\rm
D}$ and $\dot M_{\rm R}$ can vary more or less independently, and both
contribute to the total luminosity ($L_{\rm D} + L_{\rm R}$), the
source traces out separate tracks in the $\nu - I_{\rm X}$ diagram.
(The argument is more general, and is still valid if $L_{\rm R}$ is
produced by another source of energy, not necessarily related to a mass
inflow.) None of the kHz QPO models so far discussed explicitly
predicted the existence of the parallel track phenomenon, and there are
therefore no specific predictions for how QPO amplitude varies across
tracks. For this reason in this paper we use as a reference the
simplest possible scenario, namely, one in which the additional source
of energy that produces the extra luminosity producing the parallel
tracks does not affect the QPO amplitudes at all. In such a scenario,
at a constant kHz QPO frequency (i.e., constant $\dot M_{\rm D}$), the
rms amplitude of the QPOs should decrease with count rate, as the
source moves from one track to another in the $\nu - I_{\rm X}$
diagram.

In this paper we use RXTE data to study the rms fractional amplitude of
the kHz QPOs on different tracks in the $\nu - I_{\rm X}$ diagram, that
is, as a function of count rate for a fixed QPO frequency. In
\S\ref{observations} we describe the observations and results, and in
\S\ref{discussion} we discuss the significance of our findings within
the scenario previously described.

\section{Observations and Results} \label{observations}

We used observations from the Proportional Counter Array (PCA) on board
RXTE of the kHz QPO sources 4U 1608--52, 4U 1728--34, and Aql X-1. For
4U 1608--52 the data are the same as those in \citet{berger1608} and
\citet{mendez_1608_2}, except that here we do not use the observation
of 1996 March 6 because only photons from a limited energy range were
recorded. For 4U 1728--34 we analyzed the same data as described in
\citet{strohmayer96, strohmayer97}, \citet{ford&vdk98},
\citet{mendez_1728}, and \citet{disalvo_1728}. For Aql X-1 we analyzed
the same observations as used by \citet{zhangAqlx1, cuiAqlx1}, and
\citet{reigAqlx1}. We discarded those observations for which fewer than
five detectors of the PCA were on. This means a loss of $\sim 2$\,\%,
$\sim 2$\,\%, and $\sim 11$\,\% of the observations of 4U 1728--34, 4U
1608--52, and Aql X-1, respectively. A few X-ray bursts occurred during
these observations, but we excluded those parts of the data from the
rest of our analysis.

In all these observations data were recorded using the Standard 2 mode,
with 16 s time resolution in 129 energy channels, and in all cases
another mode with a time resolution of at least 1/4096 s was running in
parallel. We used the Standard 2 data to produce background subtracted
light curves (using Pcabackest v2.1b) for the full energy range of the
PCA (nominally $2 - 60$ keV).

For each source we used the high-time resolution data to produce power
spectra every 64 s, up to a Nyquist frequency of 2048 Hz, with no
energy selection. We searched these power spectra for kHz QPOs, at
frequencies $\simmore 250$ Hz. To do this we first produced a dynamic
power spectrum of each observation, we identified those power spectra
that showed a strong QPO, and we visually measured its frequency,
$\nu_{0}$. If two QPOs were present in a power spectrum, we always
picked the one at lower frequencies, because it was generally the most
conspicuous one. We fitted those power spectra, in the range $\nu_{0} -
100$ Hz to $\nu_{0} + 100$ Hz, using a function consisting of a
constant plus one Lorentzian. In a few cases in which the QPOs were not
significant enough (less than $3\sigma$, single trial) in the 64 s
segments we averaged several contiguous 64 s segments, but always less
than 20, to try and measure the QPO. We did not combine more than 20
consecutive power spectra because the QPO frequencies can vary by a few
tens of Hz within a few hundred seconds, and these frequency variations
result in biased measurements of the QPO frequencies. We discarded
those data for which this procedure did not reveal any significant kHz
QPO. We note that in this manner we might have discarded data with weak
QPOs that could have been detected using more data.

On several occasions we also detected the upper kHz QPO, either
directly in the dynamical power spectrum, or after we averaged (or
shifted and averaged; see below) the power spectra of an observation.
In those cases, to increase the significance of this QPO and to
facilitate measuring it, we aligned the individual power spectra
according to the frequency of the lower QPO, collected them into groups
such that the frequency of this QPO did not vary by more than $10-50$
Hz, and calculated the average power spectrum for each group
\citep[see][]{mendez_1608_1}. We fitted these power spectra, in a range
of 200 Hz around the approximate frequency of the upper kHz QPO, using
a constant plus one Lorentzian.

\psfig{file=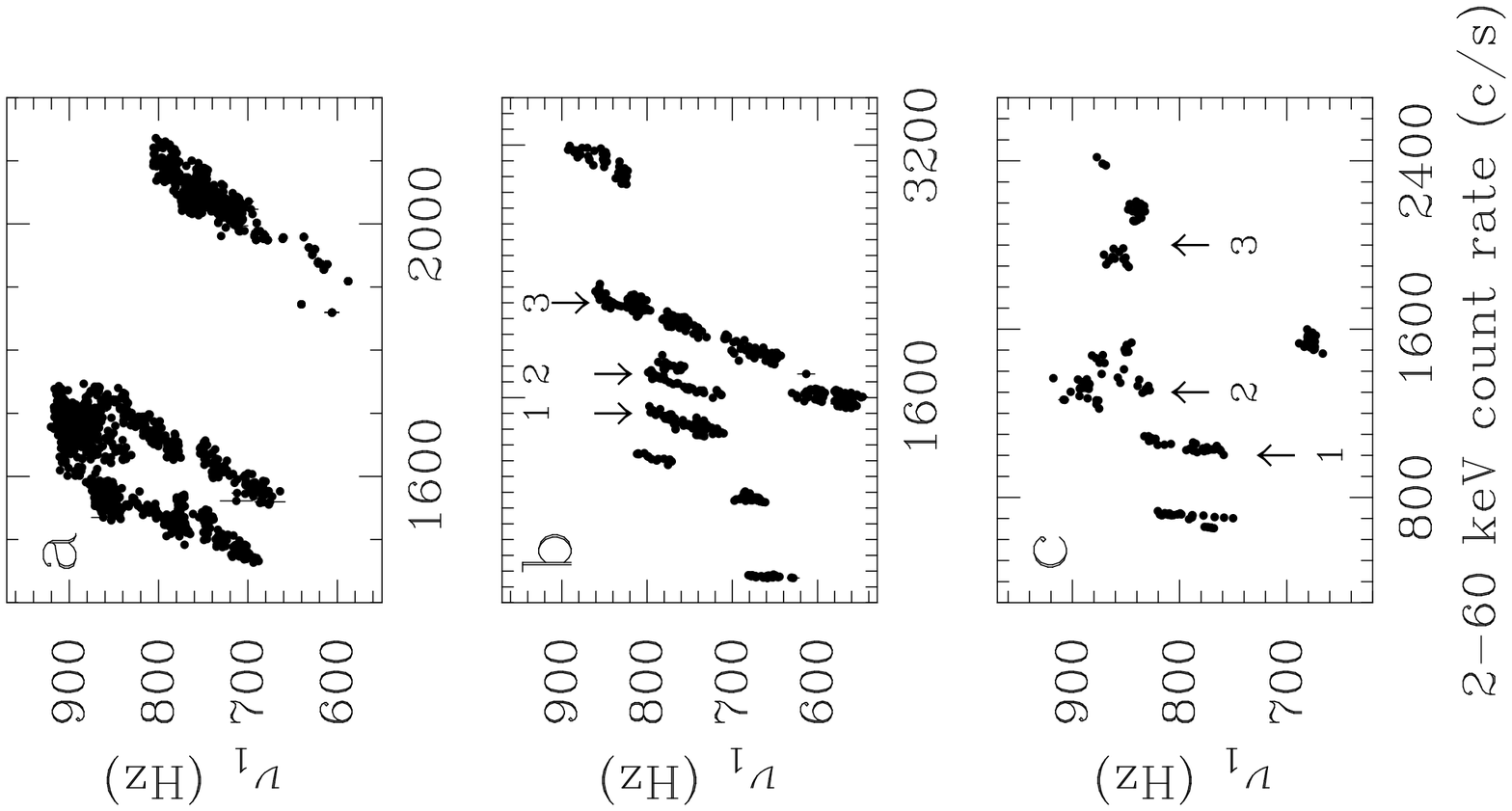,angle=-90,width=8cm}
\figcaption[fig2.ps]{The frequency of the lower kHz QPO vs. the $2 -60$
keV count rate for (a) 4U 1728--34, (b) 4U 1608--52, and (c) Aql X-1.
The count rates have been corrected for background, and are for 5
detectors. For 4U 1728--34 and 4U 1608--52 we only include data where
both kHz QPOs were detected simultaneously, so we can unambiguously
identify the lower kHz peak. For Aql X-1 we plot the only kHz QPO
detected in the power spectrum. For 4U 1608--52 and Aql X-1 we indicate
3 tracks on which we calculate the photon spectrum of the QPO (see text
and Fig. \ref{fig_rms_ene}).
\label{fig_rate}}

\vspace{0.2cm}
In those cases in which we only detected one QPO, we analyzed the
average power spectrum of that observation to try and identify it as
either the lower or the upper kHz QPO.  If the average power spectrum
still showed a single QPO, we calculated the shifted-and-averaged power
spectrum, using the frequency of the QPO we had detected as a reference
to shift the individual power spectra. If no other QPO was detected
after this procedure (this was the case for all Aql X-1 data; see
below) we treated that observation separately, so that in the rest of
our analysis the identification of the kHz QPOs as upper or lower is
unambiguous.

In Figure \ref{fig_rms_freq} we plot the rms fractional amplitude of
the QPOs that we detect in each source, as a function of their
frequency \citep[see also][Fig. 3b]{disalvo_1728}. For 4U 1728--34 and
4U 1608--52, when we detect two kHz QPOs simultaneously, we use filled
and open circles for the rms amplitude of the lower and the upper kHz
QPO, respectively; when we only detect a single QPO we use open squares
to plot its rms amplitude. For Aql X-1 we use open squares to plot the
rms amplitude of the only kHz QPO so far detected. It is apparent that
in 4U 1728--34 and 4U 1608--52 the amplitude of the upper kHz QPO is
usually larger than that of the lower kHz QPO, and that the amplitude
of the upper kHz QPO decreases monotonically with frequency, whereas
the rms amplitude of lower kHz QPO first increases, and then decreases
with frequency. When only one QPO is detected in 4U 1728--34 and 4U
1608--52, the rms amplitude of that QPO (open squares) is consistent
with the relation for the upper kHz QPO in those sources.

Aql X-1 is different from 4U 1608--52 and 4U 1728--34 in that it only
shows a single kHz QPO in its power spectrum  \citep{zhangAqlx1,
cuiAqlx1, reigAqlx1}. We applied the procedure described above to each
of the observations of Aql X-1 in which this kHz QPO was detected, but
in none of them could we detect a second QPO. Because in Aql X-1 we
only see a single kHz QPO, it is difficult to tell whether in all cases
it is the same peak, and if so, whether it is the lower or the upper
kHz QPO. However, (i) plots of the QPO frequency vs. X-ray color
\citep{mendez_texas}, or vs. position of the source in a color-color
diagram \citep{reigAqlx1} are consistent with one single relation; (ii)
the average FWHM of the QPO in Aql X-1 is $\sim 10$ Hz (it goes from
$\sim 2$ to $\sim 20$ Hz), similar to the FWHM of the lower kHz QPO
(but significantly less than the average FWHM of the upper kHz QPO) in
4U 1728--34, 4U 1608--52, and 4U 1636-53 \citep*{strohmayer96,
mendez_1608_2, mendez_1636, kaaret_1636}; (iii) the relation of the rms
amplitude of the QPO vs. the QPO frequency is similar to that of the
lower kHz QPO in 4U 1608--52 and 4U 1728--34 (see Fig.
\ref{fig_rms_freq}); (iv) the photon energy spectrum of the kHz QPO in
Aql X-1 (see below) is consistent with that of the lower kHz QPO in 4U
1608--52, and significantly different from that of the upper kHz QPO in
the same source \citep{berger1608, mendez_1608_1}. We conclude that
whenever a kHz QPO was detected in these observations of Aql X-1, it
was always the same one, and we identify it as the lower kHz QPO. In
what follows we use all the data with the kHz QPO from this source.

Figure \ref{fig_rate} shows, for 4U 1728--34, 4U 1608--52, and Aql X-1,
the relation between the frequency of the lower kHz QPO, $\nu_{1}$, and
the $2 - 60$ keV background subtracted count rate. As we explained
above, in a few occasions we had to average several consecutive 64 s
segments to get a significant measurement of the QPOs, and for that
reason each point in this figure represents between 64 ($\sim 63$\,\%
of the points) and 1280 s of data (only $\sim 8$\,\% represent more
than 256 s of data). The figure shows, for each source, the well-known
parallel tracks in the $\nu - I_{\rm X}$ diagram
\citep[e.g.,][]{ford0614_2, mendez_1608_3}.

For each source, we measured the rms fractional amplitude of the lower
and the upper kHz QPO, at constant QPO frequencies, as a function of
count rate. For this purpose we grouped the power spectra according to
the frequency of the lower QPO (Fig. \ref{fig_rate}), such that
$\nu_{1}$ did not vary by more than $\sim 40$ Hz in each frequency
range. For every source, we averaged the power spectra in each
frequency range on each of the more or less parallel tracks in Figure
\ref{fig_rate}. For every frequency and count rate interval, the
average power spectrum was calculated by first aligning the individual
power spectra according to the frequency of the lower QPO displayed in
Figure \ref{fig_rate}. In this way, if there was a second QPO in the
power spectra, it would become more significant after the alignment,
and would be easier to measure its

\begin{figure*}
\centerline{\psfig{file=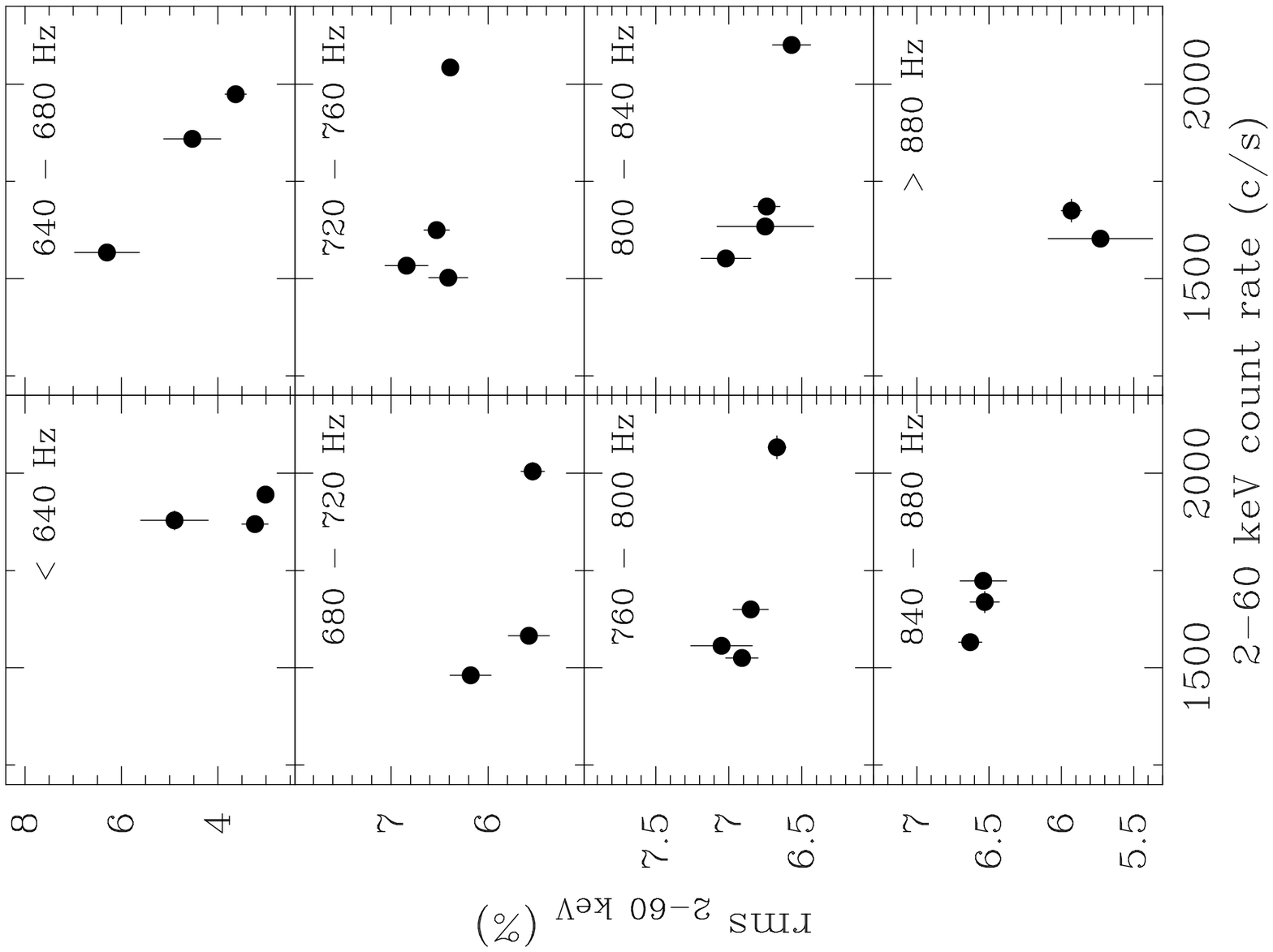,angle=-90,width=8cm}
\qquad
\psfig{file=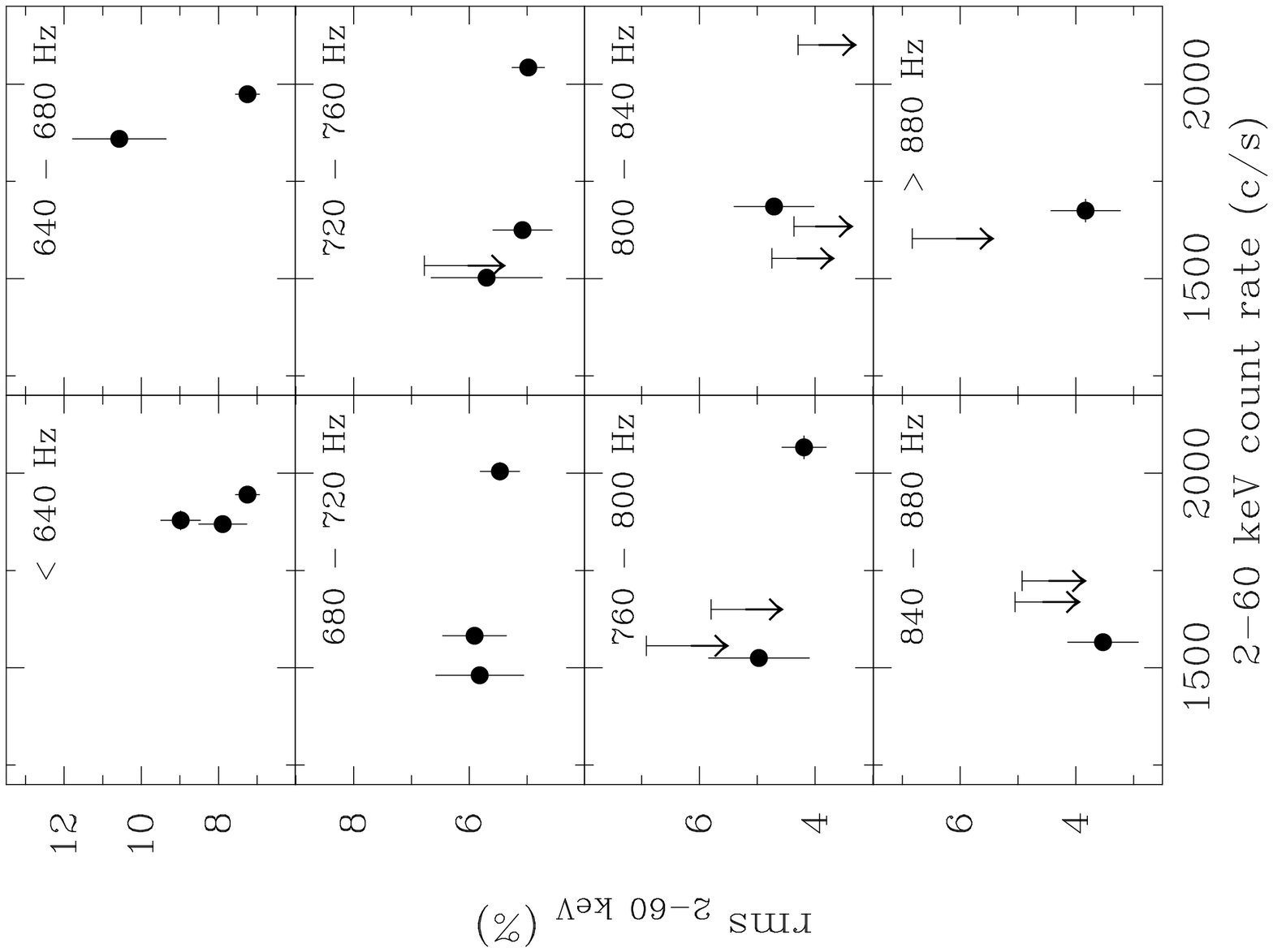,angle=-90,width=8cm}}
\figcaption[fig3.ps]{{\em a.} Rms amplitude of the lower kHz QPO in 4U
1728--34 as a function of the $2-60$ source count rate, for different
lower QPO frequency intervals as indicated. The rms amplitudes are for
the full PCA energy band. {\em b.} Same as Fig. 3a, for the upper kHz
QPO in 4U 1728--34. \label{fig_rms_1728}}
\end{figure*}

\noindent
properties \citep[cf.][]{mendez_elounda}.

We fitted each of these average power spectra (one per frequency
interval and per track for each source) using a constant plus one or
two Lorentzians. The errors in the fit parameters were calculated using
$\Delta \chi^{2} = 1$ ($1 \sigma$ for a single parameter). For 4U
1608--52 and 4U 1728--34, when the upper QPO was not significant, we
fixed its frequency to $\nu_{1} + \Delta \nu$, with $\Delta \nu$
determined from \citet[][Fig. 3]{mendez_1608_2} and \citet[][Fig.
1]{mendez_1728}, respectively, and we calculated upper limits to its
amplitude using $\Delta \chi^{2} = 2.71$, corresponding to a 95\,\%
confidence level.

\begin{figure*}[ht]
\centerline{\psfig{file=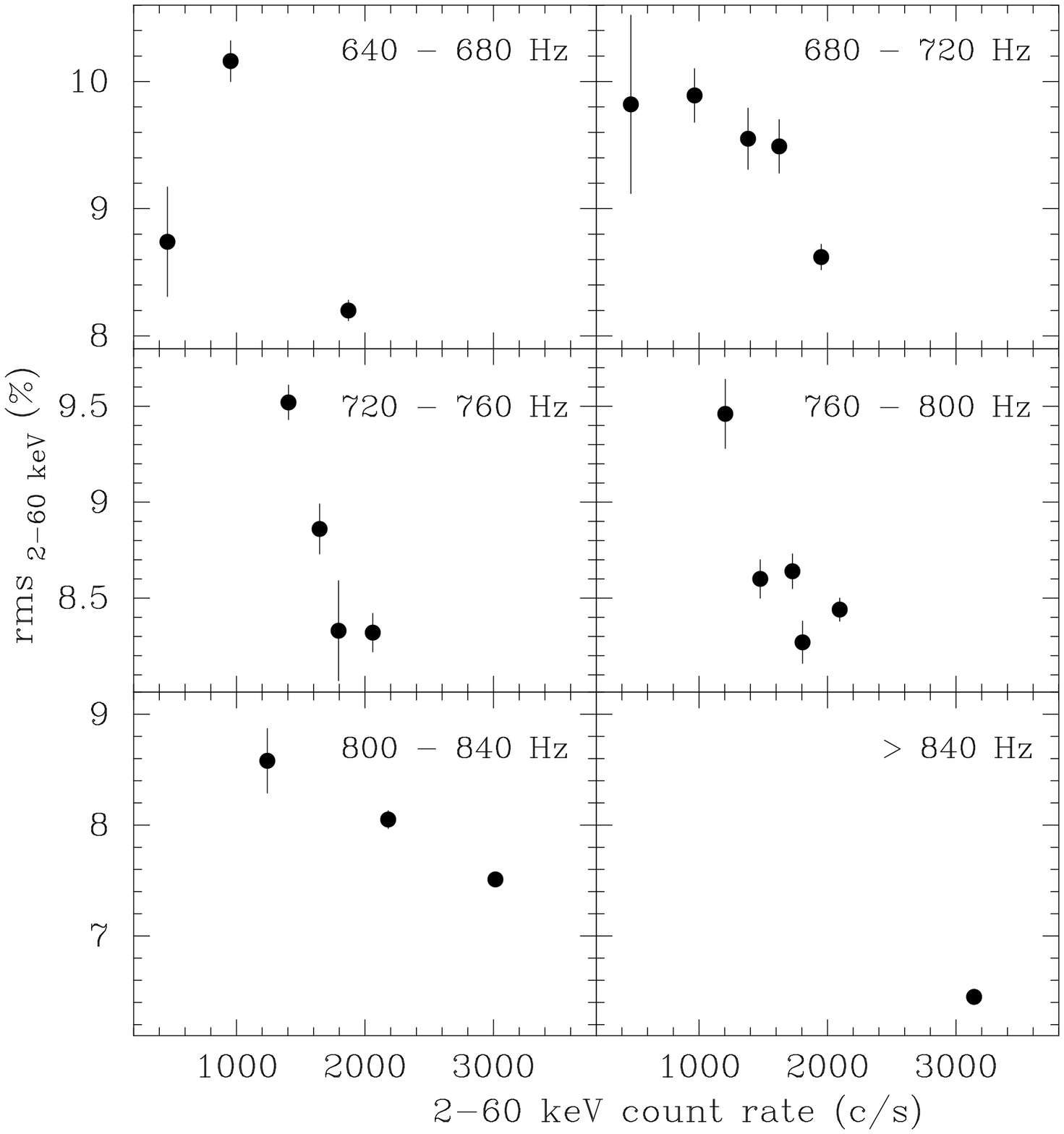,angle=-90,width=8cm}
\qquad\qquad
\psfig{file=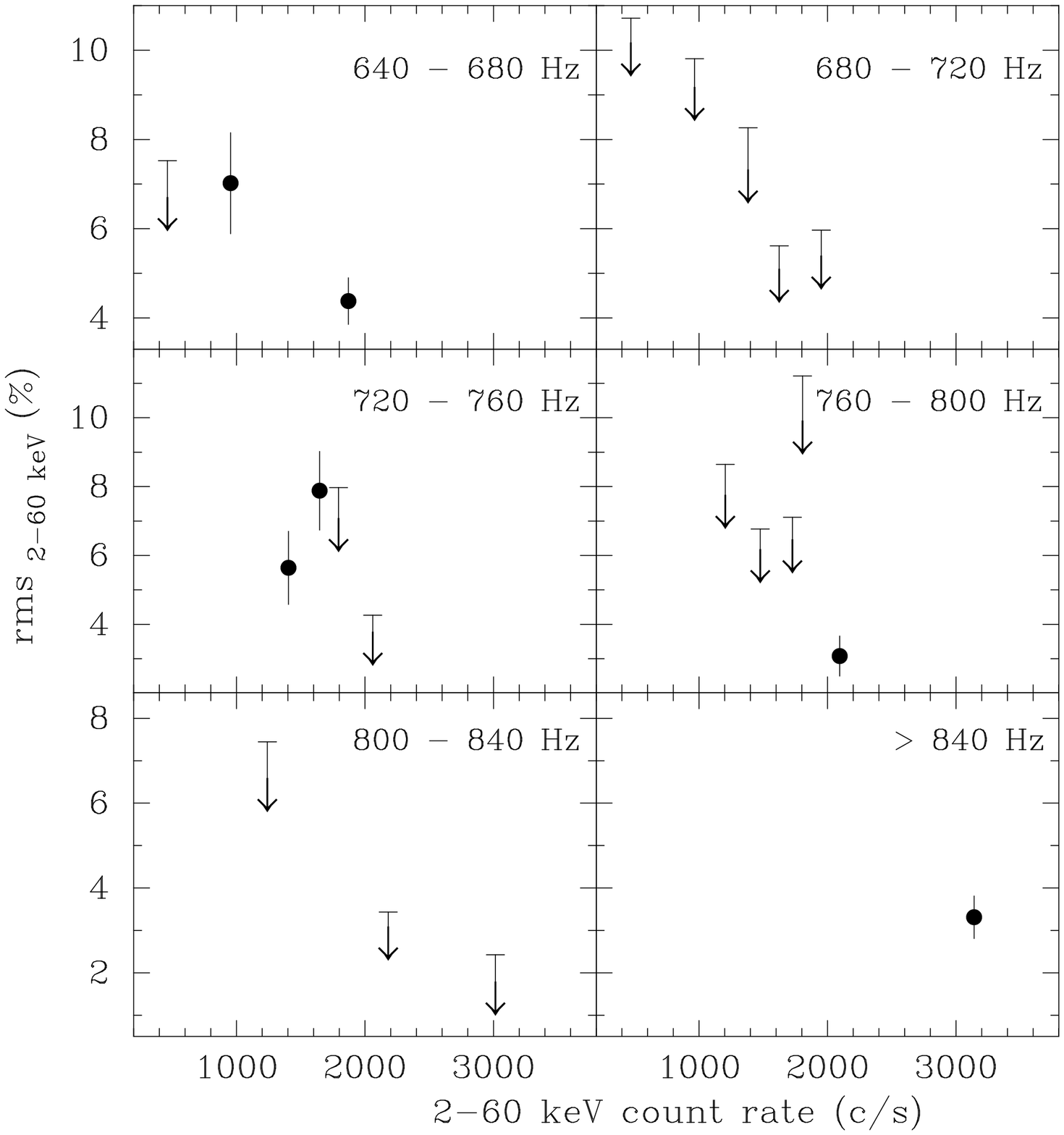,angle=-90,width=8cm}
\qquad\quad
}
\figcaption[fig4.ps]{{\em a.} Same as Fig. 3a, for the lower kHz QPO in
4U 1608--52. {\em b.} Same as Fig. 3a, for the upper kHz QPO in 4U
1608--52. \label{fig_rms_1608}}
\end{figure*}

Figures \ref{fig_rms_1728} and \ref{fig_rms_1608} show the rms
amplitude of the lower and the upper kHz QPOs as a function of count
rate, for different frequency intervals in 4U 1728--34 and 4U 1608--52,
respectively; similarly, Figure \ref{fig_rms_aqlx1} shows the rms
amplitude of the only kHz QPO in Aql X-1. Each panel in those figures
presents data for the indicated frequency interval; each point belongs
to an individual track in Figure \ref{fig_rate}. From these figures it
is apparent that, for a fixed QPO frequency, the source count rate can
increase by up to a factor of $\sim 4$, but the rms fractional
amplitude only decreases by a factor of $\sim 1.1$. There are 13 cases
for which the rms is consistent with being constant, and 9 cases in
which it decreases with count rate. 

Assuming that a fraction $\xi$ of the X-ray count rate $I_{\rm D}$
produced by the disk mass inflow $\dot M_{\rm D}$ is modulated, the
fractional rms amplitude of a QPO can be expressed as $rms = \xi I_{\rm
D} / (I_{\rm D} + I_{\rm R})$, where $I_{\rm R}$ is the X-ray count
rate produced by $\dot M_{\rm R}$. If all QPO properties are a function
of $I_{\rm D}$ only, at constant QPO frequency the ratio of the QPO rms
amplitudes on tracks $j$ and $k$ in the $\nu - I_{\rm X}$ diagram,
$rms_{\rm j} / rms_{\rm k}$, has to be equal to the inverse ratio of
count rates in those tracks, $I_{\rm X,k} / I_{\rm X,j}$. Except in the
case of the upper kHz QPO in 4U 1608--52 for which only two
measurements were available, we fitted a linear relation to the rms
amplitude ratio vs. the inverse count rate ratio, $rms_{\rm j} /
rms_{\rm k} = a I_{\rm X,k} / I_{\rm X,j} + b$, for each QPO
independently. For the lower kHz QPO in 4U 1728--34 we get $a= 0.16 \pm
0.09$ and $b = 0.83 \pm 0.07$, respectively, $\chi_{\nu}^{2} = 2$ for
27 degrees of freedom. For the upper kHz QPO in 4U 1728--34, $a= -0.41
\pm 0.41$ and $b = 1.23 \pm 0.36$, respectively, $\chi{^2}_{\nu} = 0.9$
for 7 degrees of freedom. For the lower kHz QPO in 4U 1608--52, $a=
0.25 \pm 0.06$ and $b = 0.75 \pm 0.04$, respectively, $\chi_{\nu}^{2} =
5.1$ for 27 degrees of freedom. For the only kHz QPO in Aql X-1, $a=
0.53 \pm 0.09$ and $b = 0.50 \pm 0.07$, respectively, $\chi_{\nu}^{2} =
1.1$ for 20 degrees of freedom. (Some of the fits are formally
unacceptable which could be due, for instance, to frequency-dependent
changes of the QPO rms.) Except for the upper kHz QPO in 4U 1728--34,
for which the error bars are too big to draw any conclusion, for all
the other QPOs both $a$ and $b$ are significantly different from the
expected values, $a=1$ and $b=0$.

In Figure \ref{fig_ratio} we summarize this information. We show the
ratio of the kHz QPOs rms amplitude vs. the inverse of the
corresponding count rate ratio, for points from the individual panels
in Figures \ref{fig_rms_1728}, \ref{fig_rms_1608}, and
\ref{fig_rms_aqlx1} (i.e., for a small frequency range of the lower kHz
QPO) taken in pairs (we only used data from panels that contain 2 or
more points in them). The dashed line in that figure shows the relation
expected in simple ``extra source of X-rays'' scenarios, $rms_{\rm j} /
rms_{\rm k} = I_{\rm X,k} / I_{\rm X,j}$. The solid line indicates the
best fit to the data using a straight line, $rms_{\rm j} / rms_{\rm k}
= a I_{\rm X,k} / I_{\rm X,j} + b$. The best fit parameters are $a=
0.28 \pm 0.04$ and $b = 0.73 \pm 0.03$, respectively (the uncertainties
are the $1 \sigma$ errors) with a reduced $\chi^2$ of 3 for 92 degrees
of freedom (which indicates that a single line is not a good fit~~~

\psfig{file=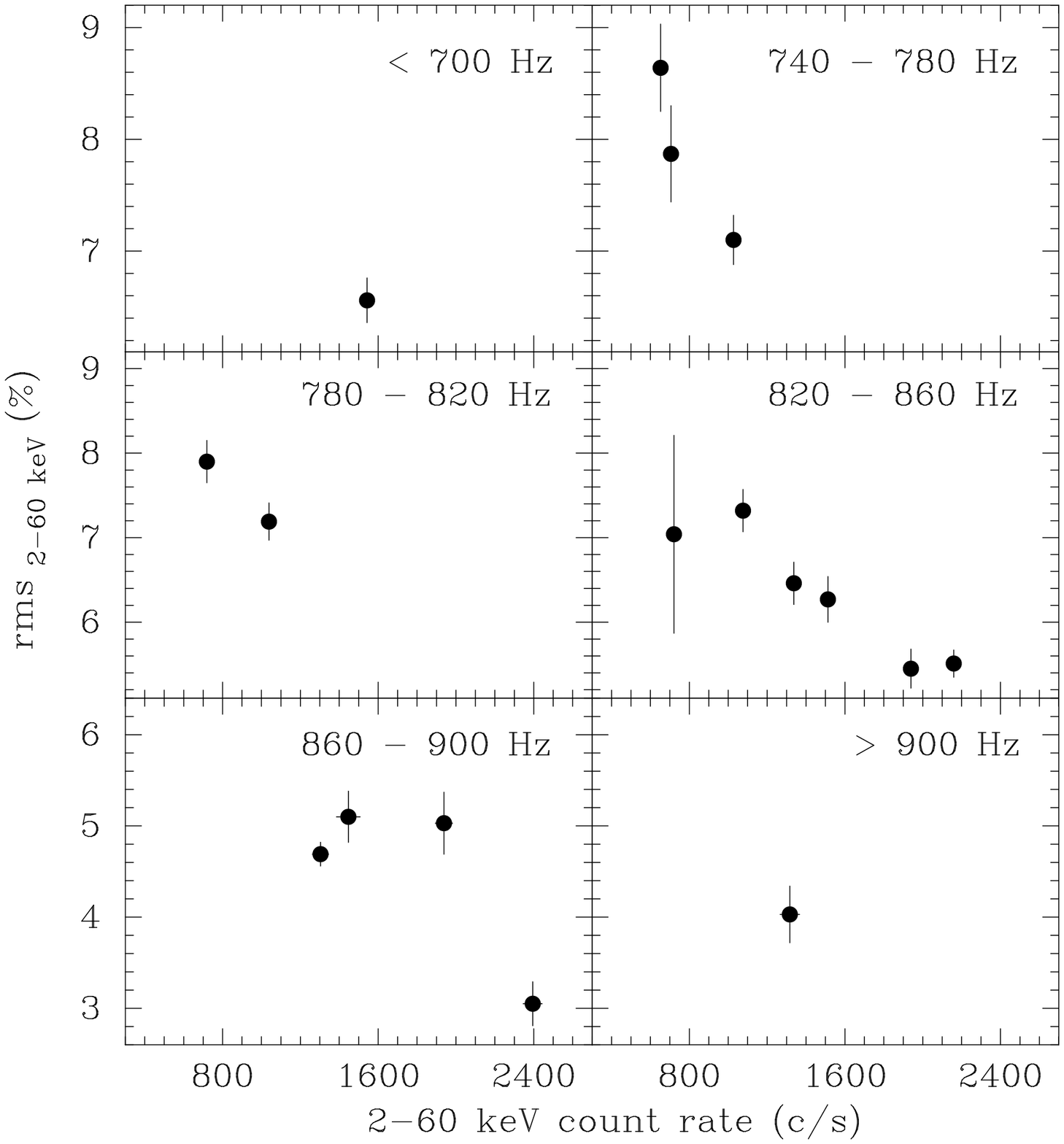,angle=-90,width=8cm}
\figcaption[fig5.ps]{Same as Fig. 3a, for the only kHz QPO in Aql X-1.
\label{fig_rms_aqlx1}}

\vspace{0.2cm}
\noindent
to the data,
as might be expected if, for example, there are some source to
source changes). We also fitted the data to a constant, which yields a
reduced $\chi^2$ of 4.8 for 93 degrees of freedom. This means an F-test
probability of $1\times 10^{-10}$ against the hypothesis that the fits
using a linear function and a constant are statistically equivalent. In
the case of the linear fit, both $a$ and $b$ are significantly
different from their expected values. The reduced $\chi^2$ is
significantly larger (54 for 94 degrees of freedom) if we fixed $a=1$
and $b=0$.

Finally, we measured the rms energy spectrum of the lower and the upper
kHz QPOs in 4U 1608--52, and the only kHz QPO in Aql X-1, when these
two sources occupied three different tracks in the $\nu - I_{\rm X}$
diagram (Fig. \ref{fig_rate}), at constant lower QPO frequency.
(Unfortunately, during the 1996 observations of 4U 1728--34 --the
points that trace out the track at the highest count rate in Fig.
\ref{fig_rate}a-- the high-time resolution data contain no energy
information, so that it was not possible to do the same analysis for
this source.) We divided the data into four energy bands, $2.0 - 4.6 -
8.0 - 13.0 - 21.5$ keV, and calculated the power spectra for both
sources in each band. In Figure \ref{fig_rms_ene} we show the rms
energy spectrum of these QPOs. For 4U 1608--52 we selected data with
QPO frequency between 700 Hz and 800 Hz, from tracks 1, 2, and 3 in
Figure \ref{fig_rate}b, at $\sim 1300 - 1500$ counts s$^{-1}$, $\sim
1600 - 1800$ counts s$^{-1}$, and $\sim 1800 - 2400$ counts s$^{-1}$,
respectively; for Aql X-1 we selected data with QPO frequency between
820 Hz and 860 Hz, from tracks 1, 2, and 3 in Figure \ref{fig_rate}c,
at $\sim 1000$ counts s$^{-1}$, $\sim 1200 - 1400$ counts s$^{-1}$, and
$\sim 1800 - 2200$ counts s$^{-1}$, respectively. For each source, we
use different symbols to indicate measurements at each of these tracks
(see figure caption). Because in 4U 1608--52 the upper kHz QPO is less
significant than the lower kHz QPO, for the upper QPO we measured the
rms spectrum using the combined data of tracks 1, 2, and 3. Figure
\ref{fig_rms_ene} shows that (i) in both sources, the rms energy
spectrum of the QPO is independent of the track that the source
occupies in the $\nu - I_{\rm X}$ diagram (Fig. \ref{fig_rate}; notice
that the overall normalization of the rms spectrum decreases as count
rate increases, corresponding to the decrease of rms amplitude with
count rate shown in Fig. \ref{fig_rms_1608} and \ref{fig_rms_aqlx1});
(ii) the rms spectrum of the only kHz QPO in Aql X-1 is similar to that
of the lower kHz QPO in 4U 1608--52, but significantly different from
that of the upper kHz QPO in 4U 1608--52.

\section{Discussion} \label{discussion}

For 4U 1728--34, 4U 1608--52, and Aql X-1 the frequencies of the kHz
QPOs trace several, almost parallel, tracks when plotted vs. X-ray
count rate (Fig. \ref{fig_rate}). We measured the rms fractional
amplitude of the kHz QPOs in these three sources at different values of
the QPO frequency, as a function of count rate. We find that, for a
fixed QPO frequency, moving from one track to another in Figure
\ref{fig_rate}, while the count rate increases by up to a factor of
$\sim 4$, the rms fractional amplitude of the QPOs only decreases by a
factor of $\sim 1.1$ (Fig. \ref{fig_rms_1728}, \ref{fig_rms_1608},
\ref{fig_rms_aqlx1}, and
\psfig{file=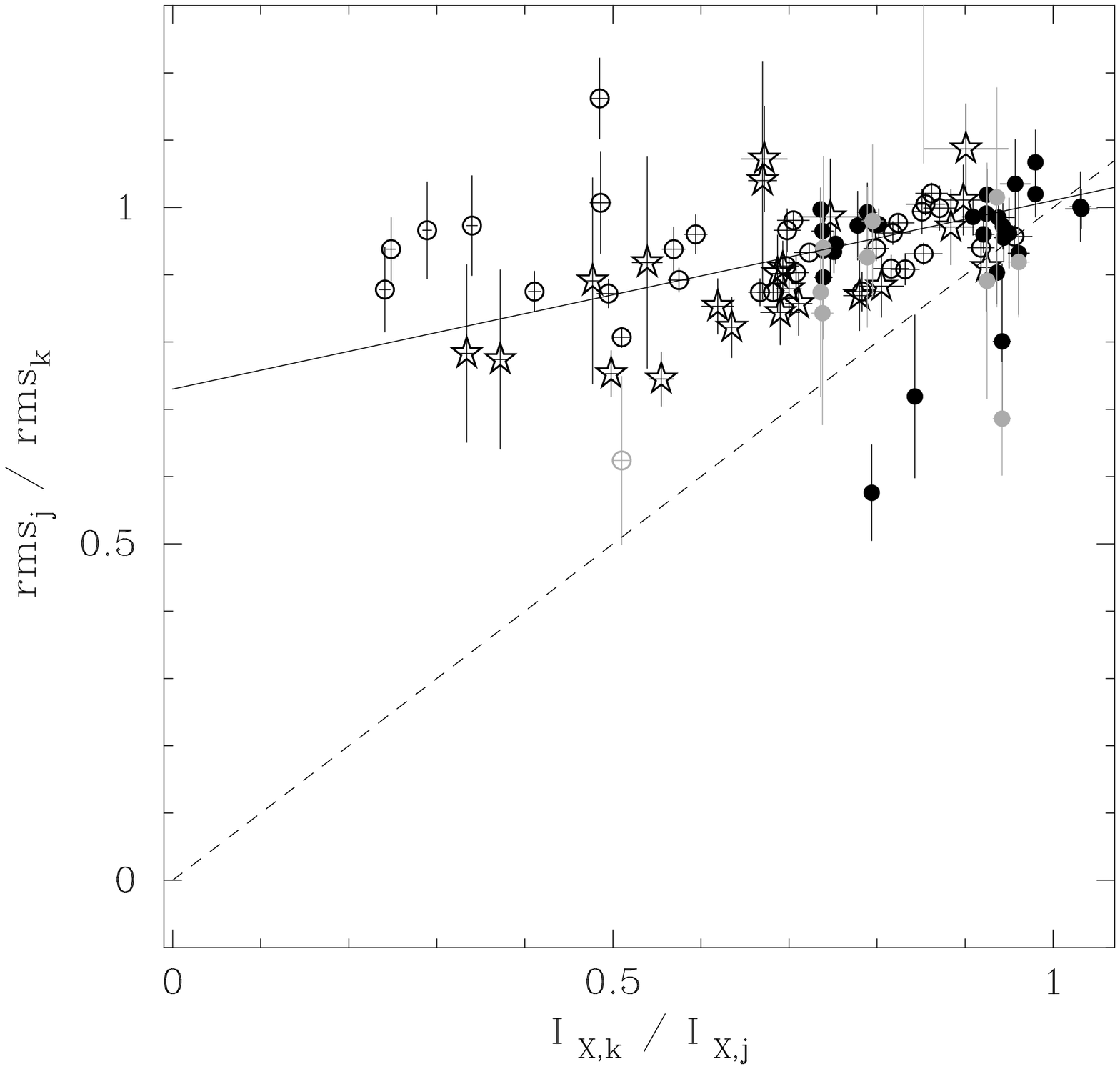,angle=-90,width=8cm}
\figcaption[fig6.ps]{Ratio of the kHz QPO rms amplitude vs. the inverse
of the corresponding count rate ratio, for each pair of points from the
panels with two or more points in Figures \ref{fig_rms_1728}, \ref{fig_rms_1608}
, and \ref{fig_rms_aqlx1}. The lower and upper kHz QPOs in 4U 1728--34 are plott
ed using black and gray filled circles,
respectively. The lower and upper kHz QPOs in 4U 1608--52 are plotted
using black and gray open circles, respectively. The kHz QPO in Aql X-1
is plotted using black open stars. The dashed line shows the expected
relation for the ``extra source of X-rays'' scenario with slope $1$ and intercep
t $0$ (see text). The solid line is the best fit straight line to the data, with
 slope $0.28 \pm 0.04$ and intercept $0.73 \pm 0.03$.
\label{fig_ratio}}
\vspace{0.2cm}
\noindent
\ref{fig_ratio}). This result is inconsistent
with simple ``extra source of X-rays'' scenarios in which the separate
tracks in Figure \ref{fig_rate} arise from a luminosity component that
does not affect the QPO. Furthermore, we also find that the rms
spectrum of the lower kHz QPO in 4U 1608--52 and that of the only kHz
QPO in Aql X-1 do not depend upon which track the source occupies in
Figure \ref{fig_rate}.

It is well known that the lack of correlation between kHz QPO
frequencies and X-ray count rate on time scales longer than a day (Fig.
\ref{fig_rate}) can coexist with a very good correlation between
frequency and position in the X-ray color-color diagram (see
\S\ref{introduction}). It was proposed that this dichotomy could be
explained in terms of two independent mass accretion rates, a disk and
a radial inflow; while both of them determine the luminosity of the
source, only the mass flow through the disk determines the frequency of
the kHz QPOs \citep{kaaret0614_1608, mendez_1608_3}. Our results for 4U
1728--34, 4U 1608--52, and Aql X-1 are inconsistent with version of
this scenario where the variations in the radial flow do not
appreciably affect the QPO amplitude: the rms fractional amplitudes of
the kHz QPOs measured at constant frequency do not decrease rapidly
enough as count rate increases (Fig. \ref{fig_ratio}).

The changes in the X-ray intensity may be due to anisotropies in the
radiation pattern. It is, however, unlikely that the radiation that is
modulated in the kHz QPO mechanism is anisotropic, as kHz QPOs have
been observed at roughly similar amplitude in sources spanning a large
range of inclinations \citep[e.g.,][]{barret_1916, homan_exo_khz}.
Perhaps, there is a large-scale redistribution of some of the emitted
energy over unobserved spectral bands. Alternatively, the missing
energy might be used to accelerate a jet. While radio emission has been
detected from Aql X-1 during X-ray outbursts, and perhaps from 4U
1728--34, it is unlikely that this explanation applies to 4U 1608--52,
as no radio emission has ever been
\psfig{file=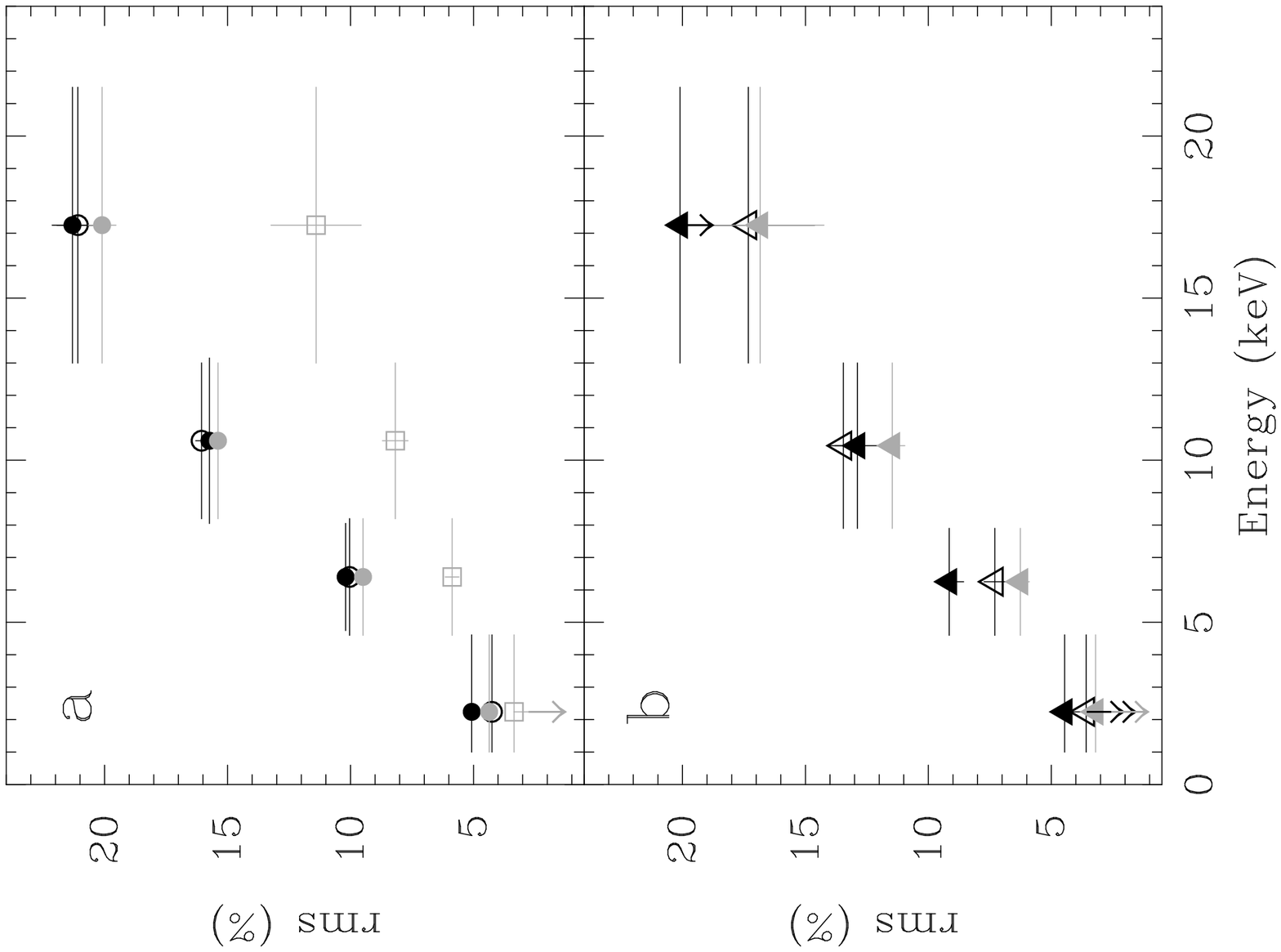,angle=-90,width=8cm}
\figcaption[fig7.ps]{Rms energy spectrum of (a) the lower and upper kHz
QPOs in 4U 1608--52, and (b) the only kHz QPO in Aql X-1. For the lower
QPO in 4U 1608--52 we use filled black, open black, and filled gray
circles to indicate measurements from tracks 1, 2, and 3 in Figure
\ref{fig_rate}b, respectively. We use gray open squares to indicate
measurements of the upper kHz QPO in 4U 1608--52. In this case we use
the combined data from tracks 1, 2, and 3. For the QPO in Aql X-1 we
use filled black, open black, and filled gray triangles to indicate
measurements from tracks 1, 2, and 3 in Figure \ref{fig_rate}c,
respectively.
\label{fig_rms_ene}}
\vspace{0.2cm}
\noindent
detected from this source
\citep{fender&hendry}.

A similar parallel tracks phenomenon is observed in the ensemble of
sources with kHz QPOs \citep*[][see also \citealt{zss97}]{vdk97,
ford_qpolx}: While for individual sources, on a time scale of hours,
QPO frequency and X-ray luminosity are in general correlation with each
other, sources that span more than two orders of magnitude in
luminosity show kHz QPOs that cover the same range of frequencies. The
fractional rms amplitudes of the kHz QPOs are $\sim 2 - 20$ times
larger in the atoll sources, at luminosities of $\sim 0.005$ to $\sim
0.2$ of the Eddington luminosity, than in the Z sources, which are
near-Eddington, and hence 5 to 400 times more luminous. For instance,
the rms amplitudes of the lower and upper kHz QPO in the Z source GX
5-1, which is thought to emit at the Eddington luminosity, are
$1-3$\,\% rms and $2-4$\,\% rms, respectively, whereas for the atoll
source 4U 0614+09, with a luminosity below $\sim 0.01$ of the Eddington
luminosity, the amplitudes of the lower and upper kHz QPOs are
$6-25$\,\%, and $6-18$\,\%, respectively \citep[][see
\citealt{hasinger89} for a definition of the Z and atoll
classes]{wijnands5-1,steve0614}. In most cases, the decrease of the
fractional rms amplitude of the kHz QPOs in the Z sources compared to
the atoll sources is less, often by large factors, than the
corresponding increase in luminosity which, as in the case of the
parallel tracks within individual sources, once again argues against
the idea that the difference in luminosity at constant QPO frequency is
only due to changes in some luminosity component that does not affect
the QPO. However, while there might be a parameter that changes from
one source to another to compensate for the dramatic increase in
luminosity leaving the QPO frequency unchanged \citep[e.g., the
magnetic field strength;][]{white97}, this can not be the explanation
for the discrepancy within individual sources discussed in this paper.

In principle, our results could be explained if the kHz QPO modulates
the luminosity from the radial inflow. However, it is unlikely that
this is the case. \citet{lee_miller} have recently suggested that it
might be possible for the Comptonizing corona to change the rms
amplitude of the kHz QPOs independently of the modulation of the
radiation emitted by the disk \cite[see also][]{miller98, lehr00}. If
this were the case, changes in the properties of the corona in
different tracks of the $\nu - I_{\rm X}$ diagram might be the reason
for the insufficiently rapid decrease of rms amplitude of the QPOs with
count rate (Fig. \ref{fig_ratio}). \citep[Although it is irrelevant to
the present issue, we note in passing that some of the conclusions in
the paper by Lee \& Miller are based on the wrong sign of the QPO phase
lags; see][]{vaughan1,vaughan2}.

However, it is known that in 4U 1608--52 (and in several other sources)
at fixed QPO frequency, the hard color does not change significantly
between different tracks in the $\nu - I_{\rm X}$ diagram
\citep[][]{kaaret0614_1608, mendez_1608_3, mendez_texas}. On the other
hand, here we have shown that the rms spectrum of the lower kHz QPO in
4U 1608--52 (and of the only kHz QPO in Aql X-1) does not depend upon
the track that the source occupies in Figure \ref{fig_rate}. As long as
the high-energy part of the X-ray spectrum is produced by inverse
Compton scattering in this corona, and the corona is also responsible
for the rms spectrum of the kHz QPO, these two results indicate that
the properties of the corona do not change significantly from one track
to another. This argues against the idea that changes in the properties
of the corona could explain the discrepancy between the observed and
the expected rate at which the rms amplitude of the kHz QPOs decrease
with count rate (Fig. \ref{fig_ratio}).

Our simple scenario does not rule out any of the QPO models so far
proposed, for none of these models explicitly predicted the parallel
track phenomenon or the behavior of the QPO rms amplitude across the
parallel tracks. But given that several of the QPO models are
compatible with the simple scenario described in this paper, our
results rejecting this scenario can be used to further constrain those
QPO models.

Recently, a new mechanism has been proposed that could explain the
parallel tracks in the $\nu-I_{\rm X}$ diagram of single sources as
well as across the ensemble of sources. Instead of two physical
parameters (e.g., $\dot M_{\rm D}$ and $\dot M_{\rm R}$) that vary
independently, \cite{vdk2001} proposes that source luminosity and kHz
QPO frequency are both determined by a single parameter, to which the
system has both an instantaneous and a long-term average response.
\cite{vdk2001} discusses scenarios where this single parameter is the
disk accretion rate $\dot M_{\rm D}$, and QPO frequency (set by the
inner disk radius) is determined by the ratio of $\dot M_{\rm D}$ to
the total luminosity. He shows that he can reproduce the observed
parallel tracks and several of their observed characteristics if the
long-term average response to $\dot M_{\rm D}$ is, for instance, $\dot
M_{\rm R}$, and luminosity depends upon the total mass accretion rate.
In models of this type it turns out that QPO frequency depends upon
$\dot M_{\rm D}$ normalized by its own long-term average (in our
example, $\dot M_{\rm R}$). So, points with the same QPO frequency on
different tracks correspond, in this scenario, to source states where
both $\dot M_{\rm D}$ and $\dot M_{\rm R}$ are different, but their
ratio is the same. The relative strengths of the QPOs in the two cases
will depend on the strengths with which they are formed (at the same
disk radius but under conditions of different accretion rate), and on
the propagation effects of the QPO signal through radial flows of
different density. This model predicts no simple linear dependence on
the amount of ``extra'' X-rays. If formation strengths are similar then
a higher-density radial flow, which (dependent on its speed) would
probably occur at higher luminosity, might suppress the QPOs more.


\acknowledgments

We would like to thank various participants of the 1999 workshop on
X-ray Probes of Relativistic Astrophysics at the Aspen Center for
Physics for pleasant and extremely fruitful discussions.  We are
specially grateful to Phil Kaaret, Fred Lamb, Coleman Miller, Dimitrios
Psaltis, Luigi Stella, and Will Zhang. MM thanks the Max-Planck
Institut f\"ur Astrophysik in Munich, for its hospitality during the
completion of part of this work. This work was supported by the
Netherlands Research School for Astronomy (NOVA), the Netherlands
Organization for Scientific Research (NWO) under contract number
614-51-002 and the NWO Spinoza grant 08-0 to E. P. J. van den Heuvel.
MM is a fellow of the Consejo Nacional de Investigaciones
Cient\'{\i}ficas y T\'ecnicas de la Rep\'ublica Argentina. This
research has made use of data obtained through the High Energy
Astrophysics Science Archive Research Center Online Service, provided
by the NASA/Goddard Space Flight Center.\par


\begin{thebibliography}{}
\bibitem[Barret et al. (1997)]{barret_1916}
        Barret, D., Olive, J. F., Boirin, L., Grindlay, J. E., Bloser,
	P. F., Chou, Y., Swank, J. H., \& Smale, A. P. 1997, \iaucirc
	6793
\bibitem[Berger et al. (1996)]{berger1608}
        Berger, M. et al. 1996, \apj, 469, L13
\bibitem[Bloser et al. (2000)]{bloser1820}
        Bloser, P. F., Grindlay, J. E., Kaaret, P., Zhang, W., Smale, A.
	P., \& Barret, D. 2000, \apj, 542, 1000
\bibitem[Boirin et al. (1999)]{boirin_1916}
	Boirin, L., Barret, D., Olive, J. F., Grindlay, J. E., \& Bloser,
	P. F. 2000, Advances in Space Research, 25, 387
\bibitem[Cui et al. (1998)]{cuiAqlx1}
        Cui, W., Barret, D., Zhang, S. N., Chen, W., Boirin, L. \&
	Swank, J. 1998, \apj, 502, L49
\bibitem[Cui (2000)]{cui2000}
        Cui, W. 2000, \apj, 534, L31
\bibitem[Campana (2000)]{campana2000}
        Campana, S. 2000, \apj, 534, L79
\bibitem[Di Salvo et al. (2001)]{disalvo_1728}
        Di Salvo, T., M\'endez, M., van der Klis, M., Ford. E. C.,
        \& Robba, N. R. 2001, \apj, 546, 1107
\bibitem[Fender \& Hendry (2000)]{fender&hendry}
        Fender, R. P., \& Hendry, M. A. 2000, \mnras, 317, 1
\bibitem[Ford et al. (1997b)]{ford0614_2}
        Ford, E., et al. 1997b, \apj, 486, L47
\bibitem[Ford et al. (1997a)]{ford0614_1}
        Ford, E., et al. 1997a, \apj, 475, L123
\bibitem[Ford \& van der Klis (1998)]{ford&vdk98}
        Ford, E. \& van der Klis, M. 1998, \apj, 506, L39
\bibitem[Ford et al. (2000a)]{ford_qpolx}
        Ford, E. C., van der Klis, M., M\'endez, M., Wijnands, R.,
        Homan, J. Jonker, P. G., \& van Paradijs, J. 2000a, \apj, 537,
	368
\bibitem[Fortner et al. (1989)Fortner, Lamb, \& Miller]{fortner89}
        Fortner,  B., Lamb, F. K., \& Miller, G. S. 1989, \nat, 342, 775
\bibitem[Ghosh \& Lamb (1979)]{ghosh79}
        Ghosh, P., \& Lamb, F. K. 1979, \apj, 234, 296
\bibitem[Hasinger \& van der Klis (1989)]{hasinger89}
        Hasinger, G., \& van der Klis, M. 1989, \aap, 225, 79
\bibitem[Homan \& van der Klis (2000)]{homan_exo_khz}
        Homan, J., \& van der Klis, M. 2000, \apj, 539, 847
\bibitem[Jonker et al. (1998)]{jonker340+0_1}
        Jonker, P. G., Wijnands, R., van der Klis, M., Psaltis, D.,
        Kuulkers, E., \& Lamb, F. K. 1998, \apj, 499, L191
\bibitem[Jonker et al. (2000)]{jonker340+0_2}
        Jonker, P. G., et al. 2000, \apj, 537, 374
\bibitem[Kaaret et al. (1998)]{kaaret0614_1608}
        Kaaret, P., Yu, W., Ford, E. C., \& Zhang, N. S. 1998, \apj, 497
        L93
\bibitem[Kaaret et al. (1999a)]{kaaret1820}
        Kaaret, P., Piraino, S., Bloser, P. F., Ford, E. C., Grindlay,
	J. E., Santangelo, A., Smale, A. P., \& Zhang, W. 1999a, \apj,
	520, L37
\bibitem[Kaaret et al. (1999b)]{kaaret_1636}
        Kaaret, P., Piraino, S., Ford, E. C., \& Santangelo, A. 1999b,
        \apj, 514, L31
\bibitem[Lee \& Miller (1998)]{lee_miller}
        Lee, H. C., \& Miller, G. S. 1998, \mnras, 299, 479
\bibitem[Lehr et al. (2000)Lehr, Wagoner, \& Wilms]{lehr00}
        Lehr, D. E., Wagoner, R. V., \& Wilms, J. 2000, ApJ, submitted
	(astro-ph/0004211)
\bibitem[Markwardt et al. (1999)Markwardt, Strohmayer, \&
        Swank]{markwardt}
        Markwardt, C. B., Strohmayer, T. E., \& Swank, J. E. 1999, \apj,
	512, L125
\bibitem[M\'endez et al. (1998c)M\'endez, van der Klis, \& van
        Paradijs]{mendez_1636}
        M\'endez, M., van der Klis, \& van Paradijs, J. 1998c, \apj,
	506, L117
\bibitem[M\'endez et al. (1998b)]{mendez_1608_1}
        M\'endez, M., et al. 1998b, \apj, 494, L65
\bibitem[M\'endez et al. (1998a)]{mendez_1608_2}
        M\'endez, M., van der Klis, M., Wijnands, R., Ford, E. C.,
        van Paradijs, J., \& Vaughan, B. A. 1998a, \apj, 505, L23
\bibitem[M\'endez \& van der Klis (1999)]{mendez_1728}
        M\'endez, M., \& van der Klis, M. 1999, \apj, 517, L51
\bibitem[M\'endez et al. (1999)]{mendez_1608_3}
        M\'endez, M.,  van der Klis, M., Ford, E. C., Wijnands, R.,
        \& van Paradijs, J. 1999, \apj, 511, L49
\bibitem[M\'endez (2000)]{mendez_texas}
        M\'endez, M. 2000, Proc. 19th Texas Symposium on Relativistic
        Astrophysics and Cosmology, ed. J. Paul, T. Montmerle, \&
        E. Aubourg (Amsterdam: Elsevier), 15/16
\bibitem[M\'endez (2001)]{mendez_elounda}
        M\'endez, M., et al. 2001, in The Neutron Star Black Hole
	Connection, ed. C. Kouveliotou, J. van Paradijs, \& J. Ventura
	(NATO ASI Ser), in press
\bibitem[Miller et al. (1998)Miller, Lamb, \& Psaltis]{miller98}
        Miller, M. C., Lamb, F. K., \& Psaltis, D. 1998, \apj, 508, 791
\bibitem[Osherovich \& Titarchuk (1999)]{osherovich99}
        Osherovich, V., \& Titarchuk, L. 1999, \apj, 522, L113
\bibitem[Psaltis et al. (1999)Psaltis, Belloni, \& van der Klis]{pbk99}
        Psaltis, D., Belloni, T., \& van der Klis, M. 1999b, \apj, 520,
	262
\bibitem[Reig et al. (2000)]{reigAqlx1}
        Reig, P., M\'endez, M., van der Klis, M., \& Ford, E. C. 2000,
	\apj, 530, 916
\bibitem[Stella \& Vietri (1998)]{stella_lt}
        Stella, L. \& Vietri, M. 1998, \apj, 492, L59
\bibitem[Stella \& Vietri (1999)]{stella99}
        Stella, L. \& Vietri, M. 1999, \prl, 82, 17
\bibitem[Strohmayer et al. (1996)]{strohmayer96}
        Strohmayer, T. E., Zhang, W., Swank, J. H., Smale, I.,
        Titarchuk, L., Day, C. \& Lee, U. 1996, \apj, 469, L9
\bibitem[Strohmayer et al. (1997)Strohmayer, Zhang, \&
        Swank]{strohmayer97}
        Strohmayer, T. E., Zhang, W., \& Swank, J. H. 1997, \apj, 487,
        L77
\bibitem[van der Klis (1995)]{vdk95}
        van der Klis, M. 1995, in X-ray Binaries, eds. W. H. G. Lewin,
	J. van Paradijs \& E. P. J. van den Heuvel (Cambridge: Cambridge
	Univ. Press), p. 252
\bibitem[van der Klis et al. (1990)]{vdk90}
        van der Klis, M., Hasinger, G., Damen, E., Penninx, W.,
        van Paradijs, J., \& Lewin, W. H. G. 1990, \apj, 360, L19
\bibitem[van der Klis (1997)]{vdk97}
        van der Klis, M. 1997, in Astronomical Time Series, ed. D. Maoz
	et al. (Dordrecht: Kluwer), 218, 121
\bibitem[van der Klis (2001)]{vdk2001}
        van der Klis, M. 2001, \apj, submitted (astro-ph/0106291)
\bibitem[van Straaten et al. (2000)]{steve0614}
        van Straaten, S., Ford, E. C., van der Klis, M., M\'endez, M.,
        \& Kaaret, P. 2000, \apj, 540, 1049
\bibitem[Vaughan et al.(1997)]{vaughan1} Vaughan, B. A. et al.
        1997, \apjl, 483, L115 
\bibitem[Vaughan et al.(1998)]{vaughan2} Vaughan, B. A. et al.
        1998, \apjl, 509, L145 
\bibitem[White \& Zhang (1997)]{white97}
        White, N. E., \& Zhang, W. 1997, \apjl, 490, L87
\bibitem[Wijnands et al. (1996)]{wijnands96}
        Wijnands, R. A. D., van Der Klis, M., Psaltis, D., Lamb, F. K.,
        Kuulkers, E., Dieters, S., van Paradijs, J., \& Lewin, W. H. G.
        1996, \apjl, 469, L5
\bibitem[Wijnands et al. (1997)]{wijnands17+2}
        Wijnands, R., et al. 1997, \apj, 490, L157
\bibitem[Wijnands et al. (1998a)]{wijnandscygx2}
        Wijnands, R., et al. 1998a, \apj, 493, L87
\bibitem[Wijnands et al. (1998b)]{wijnands5-1}
        Wijnands, R., M\'endez, M., van der Klis, M., Psaltis, D.,
        Kuulkers, E., \& Lamb, F. K. 1998b, \apj, 504, L35
\bibitem[Wijnands \& van der Klis (1999)]{wk99}
        Wijnands, R., \& van der Klis, M. 1999, \apj, 514, 939
\bibitem[Zhang et al. (1997)Zhang, Strohmayer, \& Swank]{zss97}
        Zhang, W., Strohmayer, T. E., \& Swank, J. H. 1997, \apj, 482,
	L167
\bibitem[Zhang et al. (1998)]{zhangAqlx1}
        Zhang, W., Jahoda, K., Kelley, R. L., Strohmayer, T. E., Swank,
        J. H., \& Zhang, S. N. 1998, \apj, 495, L9
\end{thebibliography}
\end{document}